\documentclass[preprint,amsmath,amssymb,aps,showkeys,showpacs]{revtex4}
\usepackage[colorlinks=true, pdfstartview=FitV, linkcolor=blue, citecolor=red, urlcolor=magenta]{hyperref}
\usepackage{graphicx}
\usepackage{latexsym}
\usepackage{amsmath}
\usepackage{amsfonts}
\usepackage{amssymb}
\usepackage{verbatim}
\usepackage{dcolumn}
\usepackage{amssymb}
\usepackage{bm}
\usepackage{float}
\usepackage{subfigure}
\usepackage[english]{babel}
\newcommand{\be}{\begin{equation}}
\newcommand{\ee}{\end{equation}}
\newcommand{\bea}{\begin{eqnarray}}
\newcommand{\eea}{\end{eqnarray}}

\begin{document}

\newcommand{\NITK}{
\affiliation{Department of Physics, National Institute of Technology Karnataka, Surathkal, Mangaluru 575 025, India}
}

\newcommand{\IIT}{\affiliation{
Department of Physics, Indian Institute of Technology, Ropar, Rupnagar, Punjab 140 001, India
}}

\title{Rotating Black Hole with an Anisotropic Matter Field as a Particle Accelerator}

\author{Ahmed Rizwan C.L.}
\email{ahmedrizwancl@gmail.com}
\NITK
\author{Naveena Kumara A.}
\email{naviphysics@gmail.com}
\NITK
\author{Kartheek Hegde}
\email{hegde.kartheek@gmail.com}
\NITK
\author{Md Sabir Ali}
\email{alimd.sabir3@gmail.com}
\IIT
\author{Ajith K.M.}
\email{ajith@nitk.ac.in}
\NITK

\begin{abstract}
Recently, a generalised solution for Einstein equations of a rotating compact body, surrounded by matter field was proposed, which is the Kerr-Newman spacetime with an anisotropic matter. The solution possesses an additional hair, along with the conventional mass, charge and spin, which arises from the negative radial pressure of the anisotropic matter. In this article we show that, this new class of black holes can act as a particle accelerator during the collision of two generic particles in its gravitational field in the ergo-region. The centre of mass energy of the particles shoots to arbitrary high value in the vicinity of event horizon for the extremal black hole. The physical conditions for the collision to take place are obtained by studying the horizon structure and circular particle motion. The results are interesting from astrophysical perspective.

\end{abstract}


\maketitle

\section{Introduction}
The recent developments in observational astrophysics ushered new interests in black hole physics, transmuting the subject from mere theoretical aspects to a realistic perspective. The first direct evidence for the existence of astrophysical black holes came through the detection of gravitational waves originating from the collision of binary black hole system \citep{Abbott:2016blz, TheLIGOScientific:2016src, Abbott:2016nmj}. On the other hand, the physical origin and mechanism of black hole shadow gained attention after it's observation by the Event Horizon Telescope \citep{Akiyama:2019bqs, Akiyama:2019cqa, Akiyama:2019fyp}. These two great discoveries opened up new windows for probing the universe through observations, and also called for a need in devising and revisiting the theoretical tools of black hole physics. An astrophysical black hole must be a rotating one as almost all the cosmic objects are characterised by definite angular momentum. Furthermore, in a realistic scenario, a stellar black hole resides in the background of matters or fields, which is to say that black hole solution must be encompassed with the coexistence of matter field with it.

The spacetime geometry of a rotating black hole gets modified in the presence of matter or fields around it. Isotropic fluids have been studied substantially in gravity theories for a long time, whereas anisotropic matter fields gained considerable attention recently. However, a plethora of studies with anisotropic matter field have appeared in the context of compact stars, relativistic stellar
objects, self-gravitating systems, stellar objects constituting quark stars, and black holes etc. Recently a simple anisotropic matter field was introduced to a static spherically symmetric black hole solution \cite{Cho:2017nhx}. Soon a rotating case was proposed with details of thermodynamics and energy extraction process \cite{Kim:2019hfp}. The rotating solution is obtained from static solution by using the Newman-Janis algorithm, which is a generalization of Kerr-Newman spacetime to incorporate an anisotropic matter. The new solution has an additional hair which arises from the negative radial pressure of the anisotropic matter, in addition to the mass, charge and spin. The properties of the solution has considerable deviation from Kerr and Kerr-Newman solutions due to the density and anisotropy of the sorrounding matter field, which are described by the parameters $K$ and $w$, respectively. As the presence of fields or fluids modify the spacetime geometry of the black hole, the corresponding effect is expected to change the shape of the black hole as seen by a distant observer. In subsequent studies, the shadow produced by rotating black holes anisotropic matter field is investigated and a deviation from the Kerr-Newmann spacetime is reported \citep{Badia:2020pnh}. The observational prospects of these effects are also discussed in the same article.

An interesting aspect of rotating black hole is the mechanical energy extraction from it. The good old idea of harnessing rotational energy of a black hole was originally proposed by Penrose, termed as Penrose process \cite{Penrose:1971uk}. The outgoing particle carries more energy than the ingoing one, the surplus energy comes from black hole spin, under appropriate physical conditions. The efficiency of this process is high in a scenario where two particles collide in the vicinity of the black hole horizon, resulting in two or more product particles. One of the product particle will escape under suitable initial conditions carrying more energy than the ingoing ones, this process is called collisional Penrose process \cite{Piran1977}. These processes are possible only in spacetimes of rotating black holes, as they possess a region called the “ergosphere”, where particles experience a frame dragging effect. A great interest on this old idea of studying particle collision in the ergo region spurred after the proposal of the celebrated BSW mechanism \cite{Banados:2009pr}. It is shown that a rotating black hole can act as a particle accelerator, which appeared as a possible candidate for highly energetic astrophysical phenomena like active galactic nuclei, gammaray bursts and ultrahigh-energy cosmic rays. With the inclusion of BSW mechanism the efficiency of energy extraction was enhanced to a greater extent, reaching the Planck scale physics  \cite{Bejger:2012yb, Schnittman:2014zsa, Berti:2014lva, Leiderschneider:2015kwa}. In BSW mechanism, the centre-of-mass energy of particles is arbitrarily high in the vicinity of the horizon of a maximally spinning black hole, when one of the particle is approaching with critical angular momentum. As the mechanism is interesting not only from theoretical aspect but also astrophysical observational perspective, several studies soon followed  \cite{Berti:2009bk, Banados:2010kn, Jacobson:2009zg, Zaslavskii:2010aw, Wei:2010gq,Harada:2014vka, Lake:2010bq, Wei:2010vca,Liu:2010ja, Mao:2010di, Zhu:2011ae, Zaslavskii:2012fh, Zaslavskii:2012qy, Zaslavskii:2010pw, Grib:2010xj, Harada:2011xz, Liu:2011wv, Patil:2010nt, Patil:2011aw, Patil:2011ya, Patil:2011uf, Amir:2015pja, Ghosh:2014mea, NaveenaKumara:2020rmi}. In this article we aim to study particle collision in the neighbourhood of a rotating black hole surrounded by an anisotropic matter field. The primary motivation for this study comes from the observation that this new solution allows the black hole to have spin more than that of Kerr and Kerr-Newman solutions which will have enhancing effects on the rotational energy extraction. Besides this, the present solution sets a more realistic set-up  for an astrophysical black hole, which has observational importance. This in turn is related to the modification in the horizon structure of the black hole, hence influences the event horizon size and static limit surface which are sought for. As we will see, the parameters related to the anisotropic matter field has significant effect on the BSW mechanism.

The article is organised as follows. In the next section (\ref{secone}) we discuss the properties of the black hole, mainly its horizon structure. In section \ref{sectwo}, we study the particle motion in the vicinity of black hole by solving the equations of motion. In section \ref{secthree}, the properties of the centre-of-mass energy of two general particles are studied, for both the extremal and non-extremal cases. We present our findings and discussions in section \ref{secfour}.


\section{Rotating black hole with an anisotropic matter field}
\label{secone}
In this section, we present the solution of a rotating black hole with an anisotropic matter field. We show that the the horizon structure and ergo region of the black hole are influenced by the matter field. The action that leads to the field equations corresponding to the family of rotating black hole solutions with an anisotropic matter field is  \citep{Kim:2019hfp},

\begin{equation}
    \mathcal{I}=\int d^4x\sqrt{-g} \left[ \frac{1}{16\pi } (R-F_{\mu \nu} F^{\mu \nu})+\mathcal{L}_m \right],
\end{equation}
where $R$ is the Ricci scalar, $F_{\mu \nu}$ is the electromagnetic field tensor and $\mathcal{L}_m$ is the Lagrangian density corresponding to the effective anisotropic matter fields. The term corresponding to anisotropic matter field can be a result of an extra $U(1)$ field or other diverse dark matters. The rotating solution for the action is obtained by using the Newman-Janis algorithm, which, in Boyer-Lindquist coordinates, has the form \citep{Kim:2019hfp, Badia:2020pnh},

\begin{equation}
    ds^2=-\frac{\rho ^2 \Delta}{\Sigma}dt^2+\frac{\Sigma \sin ^2\theta }{\rho ^2} (d\phi -\Omega dt)^2+\frac{\rho ^2}{\Delta} dr^2+\rho ^2d\theta ^2, 
    \label{metric}
\end{equation}
where
\begin{equation}
    \rho ^2=r^2+a^2 \cos ^2\theta 
\end{equation}
\begin{equation}
    \Delta =\rho ^2F(r,\theta) +a^2\sin ^2 \theta
    \label{deltaeqn}
\end{equation}
\begin{equation}
    \Sigma = (r^2+a^2)^2-a^2\Delta \sin ^2 \theta
\end{equation}
\begin{equation}
    \Omega =\frac{[1-F(r,\theta )] \rho ^2a}{\Sigma} 
\end{equation}
and
\begin{equation}
    F(r,\theta)=1-\frac{2Mr-Q^2+Kr^{2(1-w)}}{\rho ^2}.
\end{equation}
In the above expressions $M$ is the mass $a=J/M$ is the angular momentum per unit mass and $Q$ is the electric charge of the black hole. The parameter $K$ controls the density of the fluid surrounding the black hole, whereas, $w$ represents its anisotropy \citep{Kim:2019hfp, Cho:2017nhx, Kiselev:2002dx}. Similar black hole solution exists in literature, where the rotating case of a black hole with quintessential energy is obtained via Newman-Janis algorithm \citep{Toshmatov:2015npp}. However the presence of electromagnetic field and corresponding electric charge makes the present black hole solution distinct. The metric (\ref{metric}) reduces to Kerr-Newman solution when $K=0$. The energy conditions put constraints on the parameters $w$ and $K$. The metric is asymptotically flat for $w>0$, the case which we will be focusing on in this article. For the range of values $0\leq w\leq 1/2$ the total energy density is not localized sufficiently and hence the total energy diverges. Therefore, we consider only the cases with $w>1/2$, which gives asymptotically flat geometry. $w=1$ case corresponds to the matter field describing an extra $U(1)$. The energy conditions leads to the condition $Q^2+(1-2w)Kr^{2(1-w)}>0$, to have positive energy density at a radius $r$ in the rest frame of matter surrounding the black hole. In this article we will be considering both positive and negative values for $K$, which satisfy this condition. We focus on four different cases where $(w<1, k<0)$, $(w>1, k<0)$, $(w<1, k>0)$ and $(w>1, k>0)$ with the above constraints.

\begin{figure}[tbp]
\centering
\subfigure[][]{\includegraphics[scale=0.8]{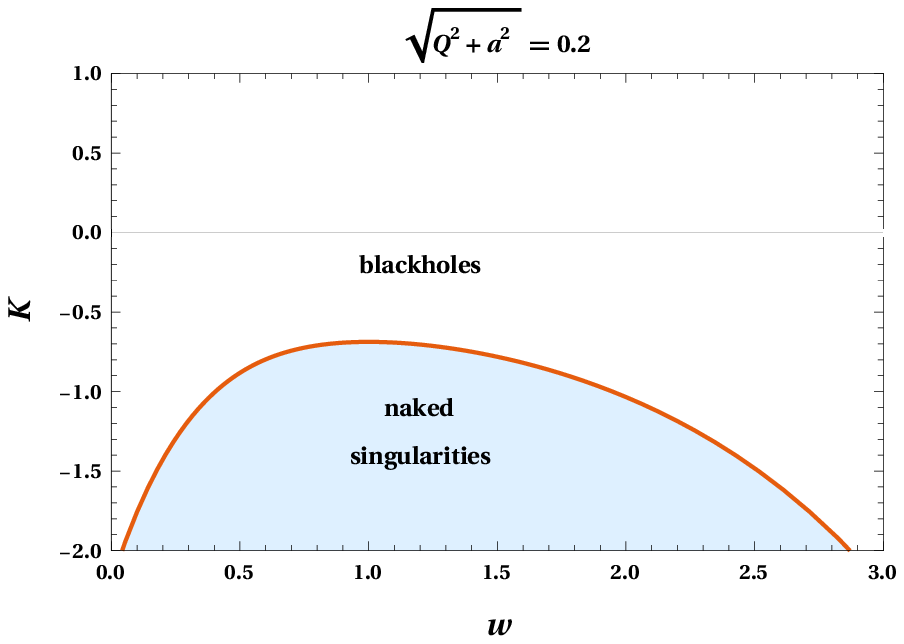}\label{horizon1}}
\subfigure[][]{\includegraphics[scale=0.8]{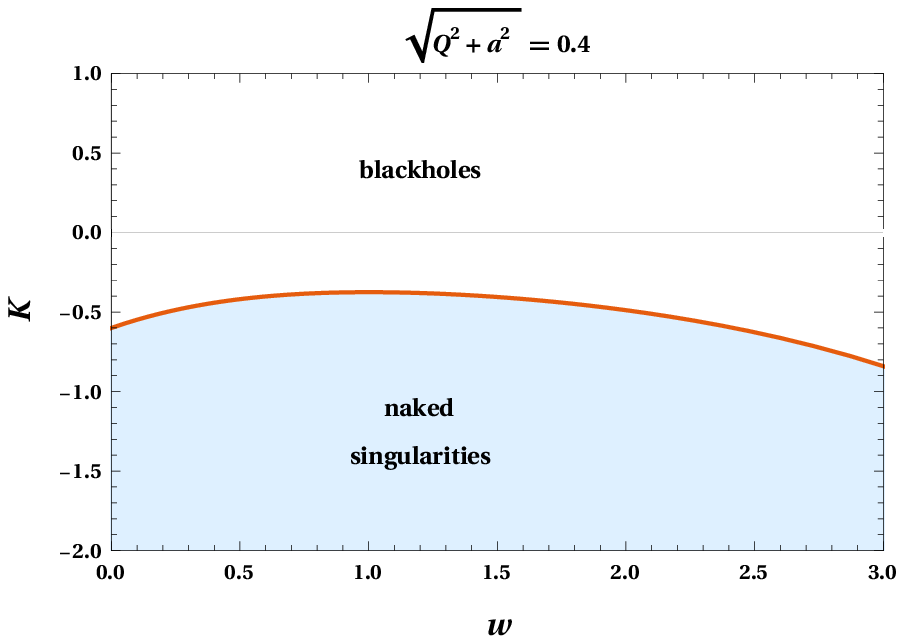}\label{horizon2}}
\caption{The parameter space of $w$ and $K$ for black hole and naked singularity solutions. The solid (orange) line is the separation curve where the event horizon disappears. We have taken the black hole mass $M=1$ in these plots.}
\label{horizondomain}
\end{figure}

The spacetime (\ref{metric}) is stationary and axisymmetric, which has Killing vectors $\partial _t$ and $\partial _\phi$ representing the time translation and rotational invariance, respectively. The metric is singular at $\Delta=0$, the largest root of which defines the event horizon of the black hole. Substituting the function $F(r,\theta)$ into Eq. \ref{deltaeqn} we have,
\begin{align}
        \Delta &=r^2+a^2+Q^2-2Mr-Kr^{2(1-w)}\\
   &=\Delta _{KN} -Kr^{2(1-w)},
\end{align}
where $\Delta _{KN}=r^2+a^2+Q^2-2Mr$ is the function  $\Delta$ found in Kerr-Newman solution. Before proceeding to the horizon structure of the black hole, we find the domains in the parameter space of $(K,w)$ for which the event horizon exists. The disappearance of event horizon is governed by the simultaneous solution of relations, $\Delta =0$ and $\Delta ' =0$. Explicitly we have,
\begin{eqnarray}
\Delta=r^2+a^2+Q^2-2Mr-Kr^{2(1-w)}=0, \\
\Delta '= 2(r-M) -2(1-w) Kr^{(1-2w)}=0.
\end{eqnarray}
Analytic solution to this simultaneous set of equations is not always feasible, hence a parametric solution is sought for as discussed in Ref \citep{Badia:2020pnh}. Rewriting first equation for $K$ and substituting that in second we get,
\begin{equation}
    w=\frac{Q^2+a^2-Mr}{\Delta _{KN}}
\end{equation}
Plugging this back into the first equation, we obtain,
\begin{equation}
    K=\frac{\Delta _{KN}}{r^{2r(r-M)/\Delta _{KN}}}
\end{equation}
The results are plotted in Fig \ref{horizondomain} for fixed $\sqrt{Q^2+a^2}$. The solid curve separates the naked singularities from black hole solutions. For a fixed value of $M$, $Q$ and $a$ we can find the allowed values of $K$ and $w$ from these plots. However, in our work, for a given values of possible $w$ and $K$, we have to numerically solve for other quantities, say, black hole spin $a$ and horizon radius $r_H$, by fixing remaining quantities, $M$ and $Q$, within physically meaningful domain, using these plots. Without loss of generality, we chose $M=1$ and $Q=0.5$ throughout this article.

\begingroup
\setlength{\tabcolsep}{8pt} 
\renewcommand{\arraystretch}{1.2} 
\begin{center}
\begin{table}
\centering
\begin{tabular}{ |c|c|c|c|c|c|c|c| } 
\hline
\multicolumn{4}{|c|}{($w =2/3$ , $k=-0.1$) }&\multicolumn{3}{|c|}{($w =3/2$ , $k=-0.1$)}\\
\hline
$a$&$r_H^+$&$r_H^-$ &$\delta ^H$ &$r_H^+$&$r_H^-$ &$\delta ^H$\\
\hline
\hline
0.1 & 1.77049 & 0.156841 & 1.61365 & 1.82782 & 0.33533 & 1.49249 \\ 
0.2 & 1.75146 & 0.17625 & 1.57522 & 1.80915 & 0.349157 & 1.45999 \\ 
0.3 & 1.71869 & 0.209612 & 1.50908 & 1.777 & 0.373622 & 1.40338 \\ 
0.4 & 1.67024 & 0.258792 & 1.41145 & 1.72951 & 0.41113 & 1.31838 \\
$a_e$&  0.966283 & 0.966283 & 0 & 1.04572 & 1.04572 & 0 \\ 
\hline
\end{tabular}
\caption{\label{cauchyandevent1} The event horizon $r_H^+$ and the Cauchy horizon $r_H^-$ for the rotating black hole with an anisotropic matter field. The values are obtained for negative value of $k$ with different $w$ values. The difference between two horizons $\delta ^H$ for non-extremal cases also shown.}
\end{table}
\end{center}
\endgroup

\begingroup
\setlength{\tabcolsep}{8pt} 
\renewcommand{\arraystretch}{1.2} 
\begin{center}
\begin{table}
\centering
\begin{tabular}{ |c|c|c|c|c|c|c|c| } 
\hline
\multicolumn{4}{|c|}{($w =2/3$ , $k=0.1$) }&\multicolumn{3}{|c|}{($w =3/2$ , $k=0.1$)}\\
\hline
$a$&$r_H^+$&$r_H^-$ &$\delta ^H$ &$r_H^+$&$r_H^-$ &$\delta ^H$\\
\hline
\hline
0.1 & 1.94652 & 0.125332 & 1.82119 &  1.89045 &- &-\\ 
0.2 & 1.93006 & 0.141428 & 1.78863 & 1.87371 & - & - \\ 
0.3 & 1.90195 & 0.168996 & 1.73295 & 1.8451 & - & -  \\ 
0.4 & 1.86099 & 0.209273 & 1.65171 & 1.8034 & - & - \\
$a_e$& 1.03297 & 1.03297 & 0 & 0.943877 & - & - \\ 
\hline
\end{tabular}
\caption{\label{cauchyandevent2} The event horizon $r_H^+$ and the Cauchy horizon $r_H^-$ for the rotating black hole with an anisotropic matter field. Here, the values are obtained for the positive value of $k$ with different $w$ values.  For $w >1$ and $k>0$ there is only one horizon for both extremal and non-extremal cases. Also, there is no degeneracy for horizon in extremal case with this condition.}
\end{table}
\end{center}
\endgroup

\begin{figure}[tbp]
\centering
\subfigure[][]{\includegraphics[scale=0.8]{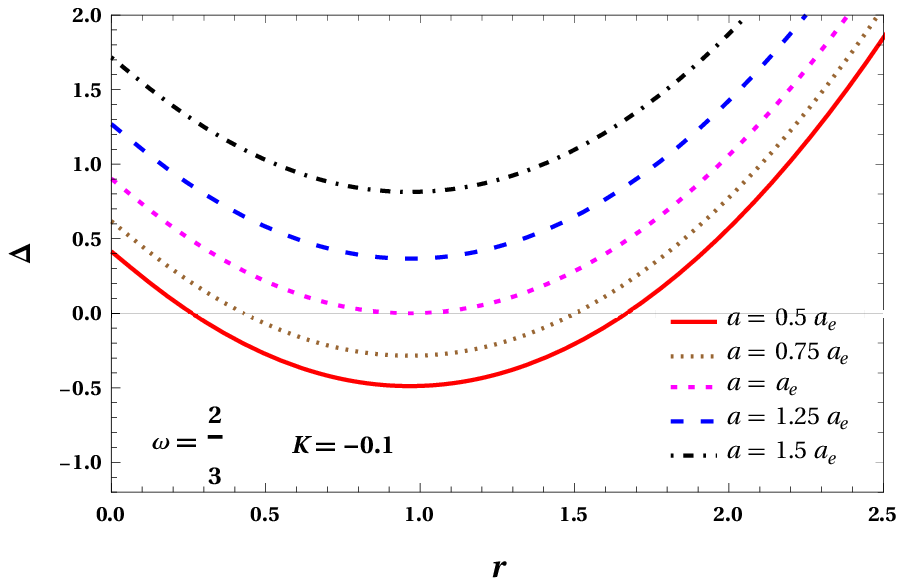}\label{kmhorizon1}}
\subfigure[][]{\includegraphics[scale=0.8]{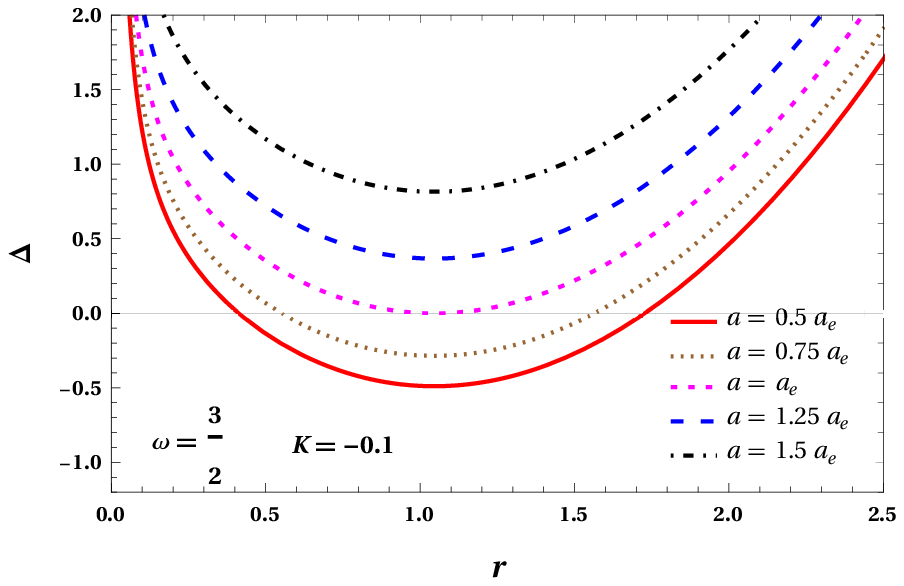}\label{kmhorizon2}}

\subfigure[][]{\includegraphics[scale=0.8]{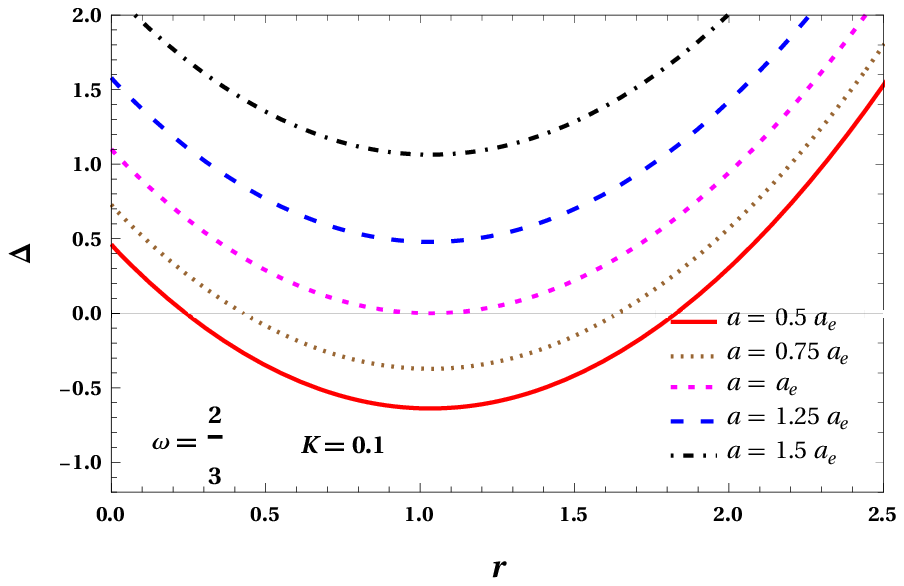}\label{kmhorizon3}}
\subfigure[][]{\includegraphics[scale=0.8]{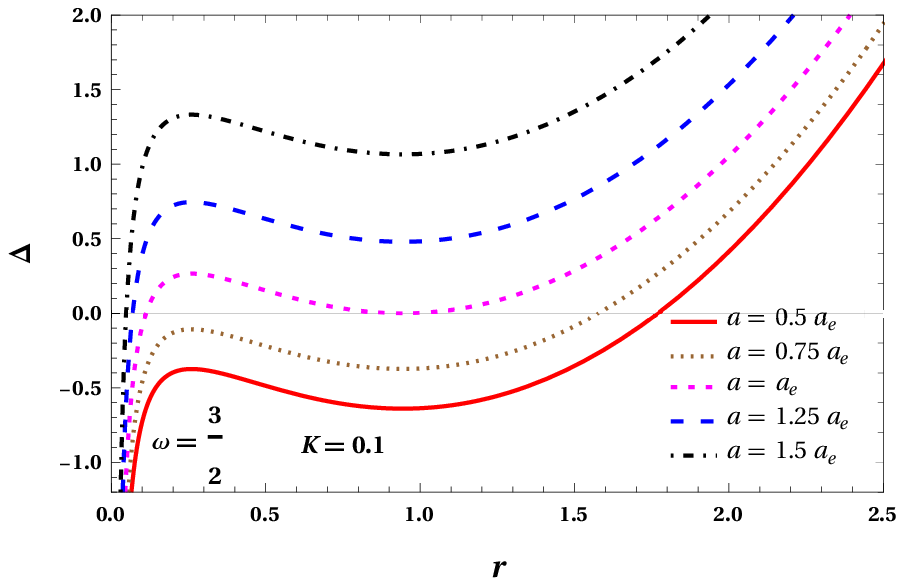}\label{kmhorizon4}}

\caption{The horizon structure of the black hole for different values of $(w,k)$. For a fixed allowed value of $(w , k)$, there exist an extremal black hole with a single horizon (represented by the dotted (Magenta) line which just touches the x-axis). For $w >1$ and $k>0$ case, the horizon structure is distinct, whereas, the extremal case horizon is degenerate. The size of the black hole changes with the choice of $(w,k)$ for a fixed mass $M$ and charge $Q$. We have taken $M=1$ and $Q=0.5$.}
\label{horizon}
\end{figure}

As in the case of Kerr metric, the solution (\ref{metric}), generally possess two horizons, namely, the Cauchy horizon and the event horizon. The event horizon is a no return surface. We study the horizon structure of the black hole for different values of allowed $(w , K)$, which is given in table \ref{cauchyandevent1}, table \ref{cauchyandevent2}. and in fig. \ref{horizon}. The difference $\delta ^H$ between the event horizon $r_H^+$ and the Cauchy horizon $r_H^-$ for different values of black hole spin $a$ are calculated. The results show that the horizon size depends on given $(w ,K)$. The spin $a_e$ corresponding to $\Delta =0 $ and $\Delta '=0$, defines the extremal black hole. The Cauchy horizon and event horizon coincide for an extremal black hole. The black holes with spin lower than $a_e$ are termed as non extremal. For cases $K>0$ and $w >1$ we note that there is no Cauchy horizon for the non extremal cases, and in the extremal case, the horizon is not degenerate.

\begin{figure}[tbp]
\centering
\subfigure[][]{\includegraphics[scale=0.8]{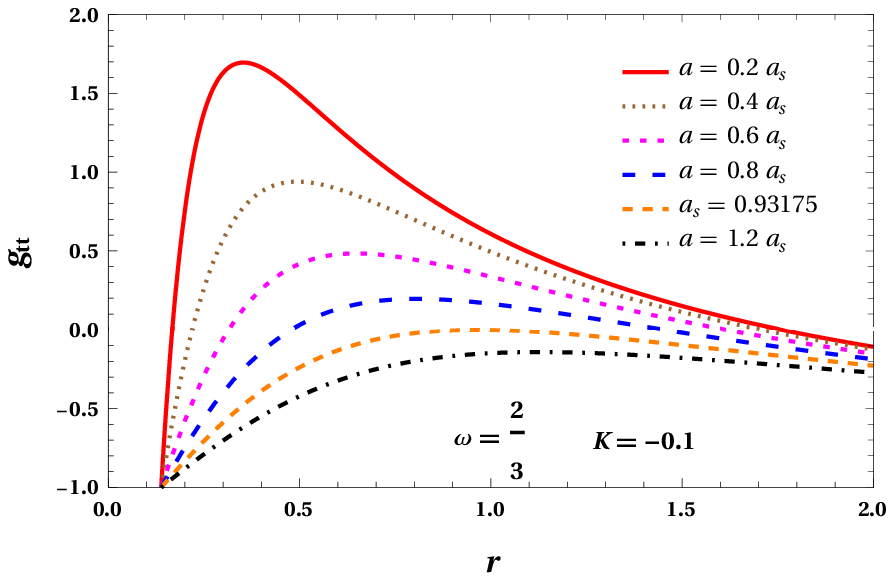}\label{kmsls1}}
\subfigure[][]{\includegraphics[scale=0.8]{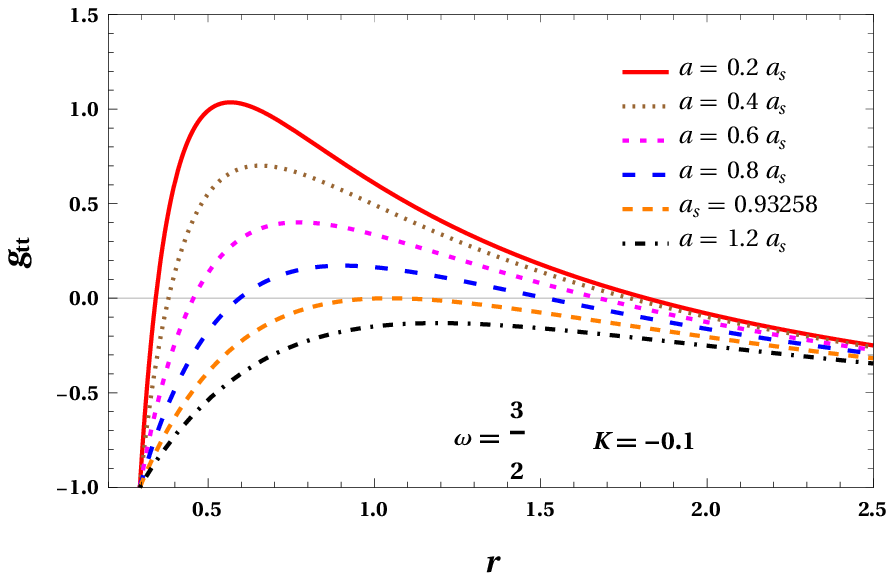}\label{kmsls2}}

\subfigure[][]{\includegraphics[scale=0.8]{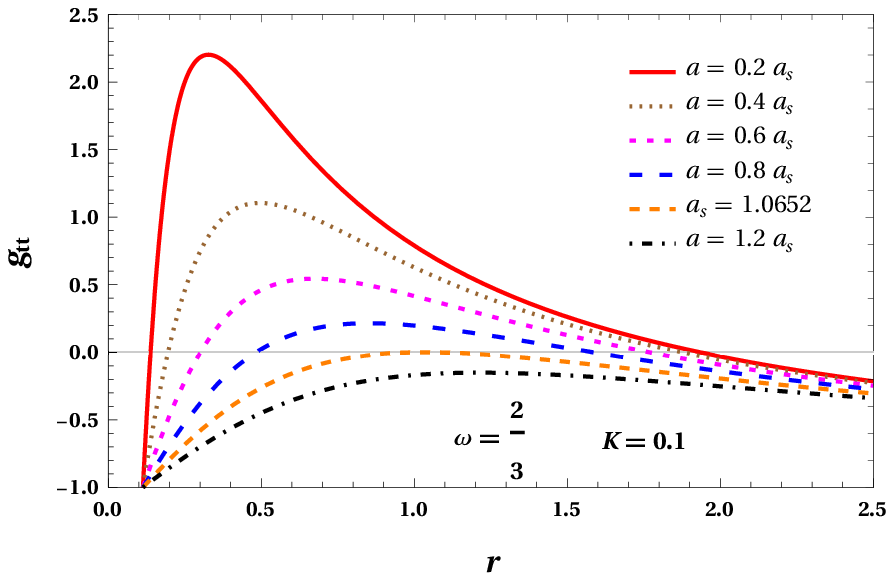}\label{kmsls3}}
\subfigure[][]{\includegraphics[scale=0.8]{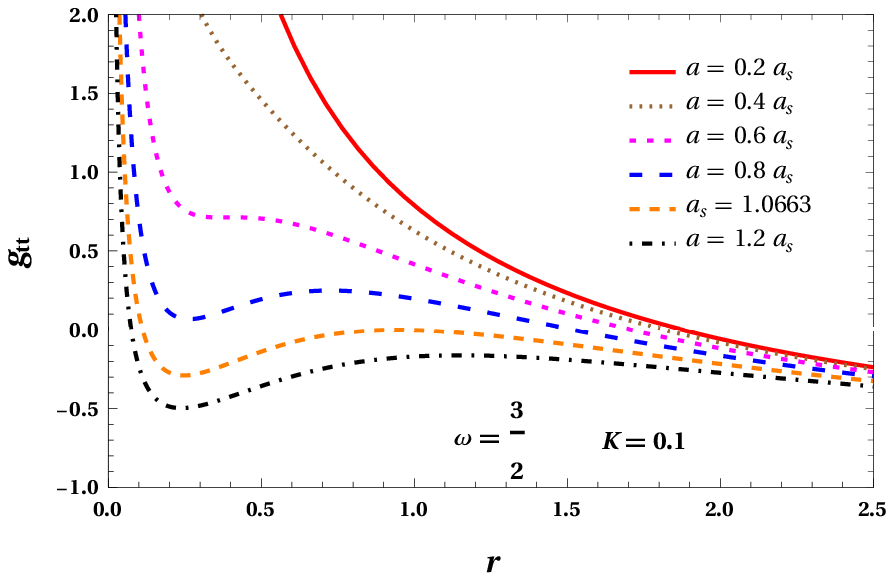}\label{kmsls4}}

\caption{The structure and location of the ergo surface of the black hole. The size of the ergo sphere depends on the choice of $(w , k)$ for a fixed value of mass $M$ and $Q$.  We have taken $M=1$ and $Q=0.5$. We have also fixed $\theta =\pi /6$. For $w >1$ and $k>0$ the ergo surface structure is distinct, as observed in the case of horizon structure.}
\label{sls}
\end{figure}

Now we focus on the interesting feature of a rotating black hole, namely the frame dragging effect. This effect leads to the creation of a \emph{no static region} outside the event horizon, called the ergo region. In the ergo sphere of the black hole, the coordinates $t$ and $r$ are spacelike. In this region, a particle is forced to move in the rotational direction of the black hole. The limiting surface of this region is static limit surface, characterised by the condition $g_{tt}=0$. Unlike event horizon, a particle crossing this surface can return and escape to infinity. The energy of the escaping particle can even be enhanced by taking a share from the rotational energy of the black hole. This makes a rotating black hole a possible candidate for a source of highly energetic astrophysical phenomena like gamma ray bursts and active galactic nuclei. The extension of the static limit surface depends on the angular coordinate  $\theta$, in addition to the black hole parameters and coincides with the event horizon at the poles. For various set of $(w , K)$ values we have studied the location of ergo region of the black hole by fixing $\theta =\pi /6$ (fig. \ref{sls}). As in the case of horizon structure, the static limit surface depends on the choice of $(w , K)$. The deviation from the common trend for the case $w>1$ and $K>1$ also present here. This shows that the mechanical energy extraction from the black hole deviates from the results obtained for Kerr-Newman black hole.

\section{Orbit of the test particle around the rotating black hole}
\label{sectwo}

In this section we study the trajectory of a test particle of mass $\mu$ and charge $q$ in a spacetime background of rotating black hole with matter field. The Lagrangian characterising this particle motion is,
\begin{equation}
    \mathcal{L}=\frac{1}{2}g_{\mu \nu} \dot{x}^{\mu} \dot{x}^{\nu} +qA_{\mu} \dot{x}^{\mu},
\end{equation}
where $A_{\mu}$ is the 4-dimensional electromagnetic potential and the dot denotes a differentiation with respect an affine parameter $\lambda$ along the geodesic. The affine parameter and the proper time $\tau$ are related to each other as $\tau =\mu \lambda$, which is equivalent to,
\begin{equation}
    g_{\mu \nu} \dot{x}^{\mu} \dot{x}^{\nu}=-\mu ^2.
\end{equation}
For uncharged particles $\mu ^2=-1,0$ and $1$ corresponds to timelike, null and spacelike geodesics, respectively. The conjugate four-momenta are, 
\begin{equation}
    P_{\mu} =\frac{\partial \mathcal{L}}{\dot{x}^{\mu}} =g_{\mu \nu} \dot{x}^{\nu} +qA_{\mu}.
    \label{pmueqn}
\end{equation}
Now, the Hamiltonian for the particle motion can be obtained as,
\begin{equation}
  H=P_{\mu} \dot{x}^{\mu} -\mathcal{L}=\frac{1}{2}g^{\mu \nu} (P_{\mu} -qA_{\mu})(P_{\nu} -qA_{\nu}).
\end{equation}
Using this Hamiltonian we write the Hamilton-Jacobi equation,
\begin{equation}
    \frac{\partial S}{\partial \lambda}=H=\frac{1}{2}g^{\mu \nu} (P_{\mu} -qA_{\mu})(P_{\nu} -qA_{\nu}),
    \label{jacobieqn}
\end{equation}
where $S$ is the Jacobi action. For the spacetime geometry of a rotating black hole with matter field, the action $S$ is separable in to a simple form,
\begin{equation}
S=\frac{1}{2}\mu  ^2\lambda -E t +L\phi +S_r(r)+S_{\theta}(\theta),
\label{jacobiaction}
\end{equation}
where $E$ is the energy of the particle and and $L$ is the azimuthal angular momentum. These quantities are constants of motion along the geodesic, which are  related to the symmetries of the spacetime, the cyclic coordinates $t$ and $\phi$, respectively, and to the associated Killing vectors. $S_r$ and $S_\theta$ are the functions of $r$ and $\theta$, respectively. Substituting Eq. \ref{jacobiaction} into Eq. \ref{jacobieqn}, we get,
\begin{equation}
    S_\theta ^2+\left(aE \sin \theta -\frac{L}{\sin \theta}\right)^2+a^2\cos ^2 \theta =-\Delta S_r^2 +\frac{\left[ (a^2+r^2)E-aL-qQr\right]^2}{\Delta}-\mu ^2  r^2
\end{equation}
The above equation is in a variable separable form, the left hand side is dependent only on $\theta$ and right hand side is only a function of $r$. Therefore, both LHS and RHS must be equal to a constant, say $\mathcal{K}$. With this, we have,
\begin{equation}
    S_\theta ^2=\mathcal{K}-\left(aE \sin \theta -\frac{L}{\sin \theta}\right)^2-a^2\cos ^2 \theta
\end{equation}
\begin{equation}
    \Delta S_r^2 =-\mathcal{K}+\frac{\left[ (a^2+r^2)E-aL-qQr\right]^2}{\Delta}-\mu ^2  r^2.
\end{equation}
Using the equation $P_r=\frac{\partial S}{\partial r}$ and  $P_\theta=\frac{\partial S}{\partial \theta}$ along with Eq. \ref{pmueqn}, we obtain,
\begin{equation}
    \frac{d\theta}{d\tau} =\pm \frac{\sqrt{\Theta}}{\rho ^2},
\end{equation}
\begin{equation}
    \frac{d r}{d\tau} =\pm \frac{\sqrt{\mathcal{R}}}{\rho ^2}
    \label{radial}
\end{equation}
with,
\begin{equation}
    \Theta =\mathcal{K}-(L-aE)^2-\cos ^2\theta \left( a^2(\mu ^2-E^2)+\frac{L^2}{\sin ^2\theta}\right),
\end{equation}
\begin{equation}
    \mathcal{R}(r)=P(r)^2-\Delta[(L-aE)^2+\mu ^2r^2+\mathcal{Q}],
\end{equation}
\begin{equation}
    P(r)=E(r^2+a^2)-La-qQr,
\end{equation}
where $\mathcal{Q}$ is the Carter constant, which is related to constant $\mathcal{K}$ as $\mathcal{Q}=\mathcal{K}-(L-aE)^2$. Using the equation $P_t=\frac{\partial S}{\partial t}$ and  $P_\phi=\frac{\partial S}{\partial \phi}$ along with Eq. \ref{pmueqn}, we obtain,
\begin{equation}
    E=-(g_{tt}\dot{t}+g_{t \phi }\dot{\phi}+qA_t)
\end{equation}
\begin{equation}
    L=g_{\phi \phi }\dot{\phi}+g_{t\phi}\dot{t}+qA_{\phi}.
\end{equation}
Solving these equations we get,

\begin{equation}
 \frac{d t}{d \tau}=\frac{a(L-aE \sin ^2 \theta)}{\rho ^2}+\frac{r^2+a^2}{\rho ^2 \Delta }P(r)
\label{teqn}
\end{equation}
\begin{equation}
 \frac{d \phi}{d \tau}=\frac{(L-aE\sin ^2 \theta ) }{\rho ^2 \sin ^2\theta}+\frac{a}{\rho ^2 \Delta}P(r).
\end{equation}
This completes calculation of the equations of motion of the particle around the black hole. Now we aim to consider the particle collision in the spacetime of rotating black hole with a matter field. However, not all particles moving towards the black hole will reach the ergo region or event horizon. We must examine the range of angular momentum of the particles to exclude the particles being scattered. Before proceeding further, we would like to mention that black holes are are surrounded by relic cold dark matter density spikes. It is a widely accepted notion that cold dark matter does not interact with other matters electromagnetically. Therefore, we consider the collision of two uncharged cold dark matter particles in the spacetime of rotating black hole with matter field, i.e., $q_1=q_2=0$. For simplicity we consider the particle collision in the equatorial plane defined by $\theta=\pi/2$. In the equatorial plane we also have vanishing Carter constant, $\mathcal{Q}=0$. To examine the range of allowed angular momentum of the particle, we consider its radial motion described by Eq. \ref{radial}, which is expressed as,

\begin{equation}
\frac{1}{2}\dot{r}^2+V_{eff}=0.
\end{equation}
where the effective potential is,
\begin{equation}
V_{eff}=\frac{\left[E \left(a^2+r^2\right)-a L\right]^2-\Delta \left[\mu ^2r^2+ \left(a E-L\right)^2\right]}{2 r ^4}.
\end{equation}
The  influence of $(w ,K)$ on the particle motion comes via $\Delta$. The circular orbits are governed by the conditions,
\begin{equation}
V_{eff}=0\quad , \quad \frac{dV_{eff}}{d r}=0.
\end{equation}
Using these constraints we obtained the limiting values $L_{min}$ and  $L_{max}$ for different values of $(w , K)$ for extremal (table \ref{lrangeext}) and non-extremal (table \ref{lrangenonext}) black holes.  We also consider the case of photons, which enjoy a greater window of angular momentum range compared to the cold dark matter particles. However, in this article, we consider only the collision of massive particles.
\begingroup
\setlength{\tabcolsep}{10pt} 
\renewcommand{\arraystretch}{1.2} 
\begin{center}
\begin{table}
\centering
\begin{tabular}{ |c|c|c|c|c|c|c|c| } 
\hline
&&&\multicolumn{2}{|c|}{massive particle}&\multicolumn{2}{|c|}{photon}\\
 \cline{4-7}
$(w , k)$ & $a_e$ & $r_H^e$ &$L_2$(min)&$L_1$(max)&$L_2$(min)&$L_1$(max)\\
\hline
\hline
$(2/3, -0.1)$& 0.806922 & 0.966283 & -4.47157 & 1.96404 & -6.31639 & 1.96404 \\ 
$(3/2, -0.1)$& 0.80764 & 1.04572 & -4.58561 & 2.16163 & --6.47615 & 2.16163 \\ 
$(2/3, 0.1)$& 0.92255 & 1.03297 & -4.79787 & 2.07917 & -6.86925 & 2.07917 \\ 
$(3/2, 0.1)$& 0.92347 & 0.943877 & -4.68799 & 1.88821 & -6.7161 & 1.88821 \\ 
\hline
\end{tabular}
\caption{\label{lrangeext} The range of angular momentum of the infalling particle (photon or massive particle) for the extremal rotating black hole with an anisotropic matter field.}
\end{table}
\end{center}
\endgroup

\begin{figure}[tbp]
\centering
\subfigure[][]{\includegraphics[scale=0.8]{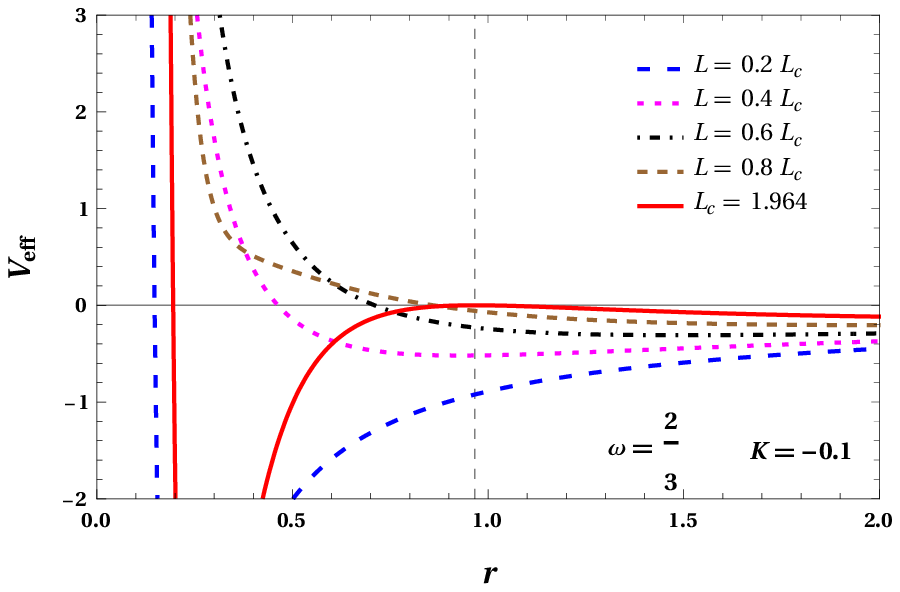}\label{kmveff1a}}
\subfigure[][]{\includegraphics[scale=0.8]{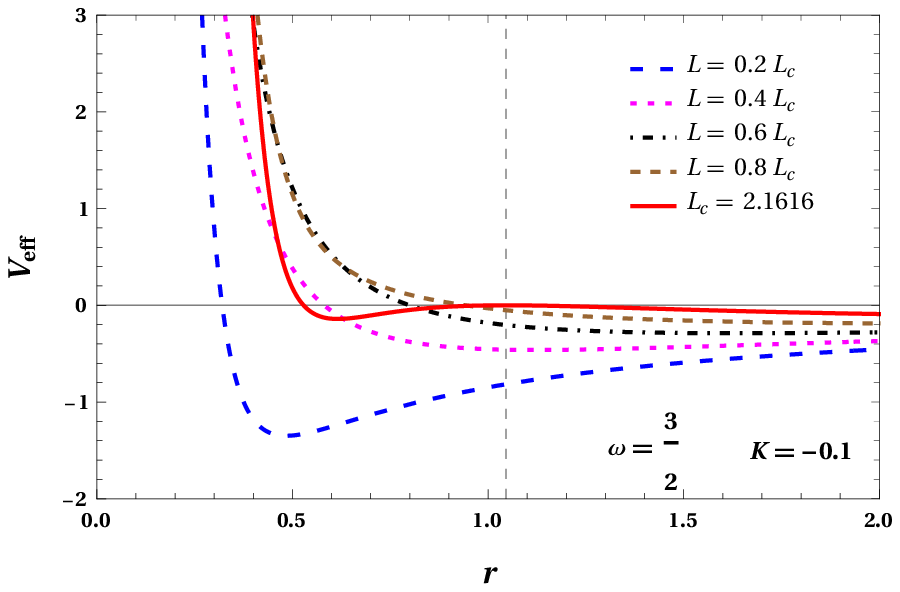}\label{kmveff2a}}

\subfigure[][]{\includegraphics[scale=0.8]{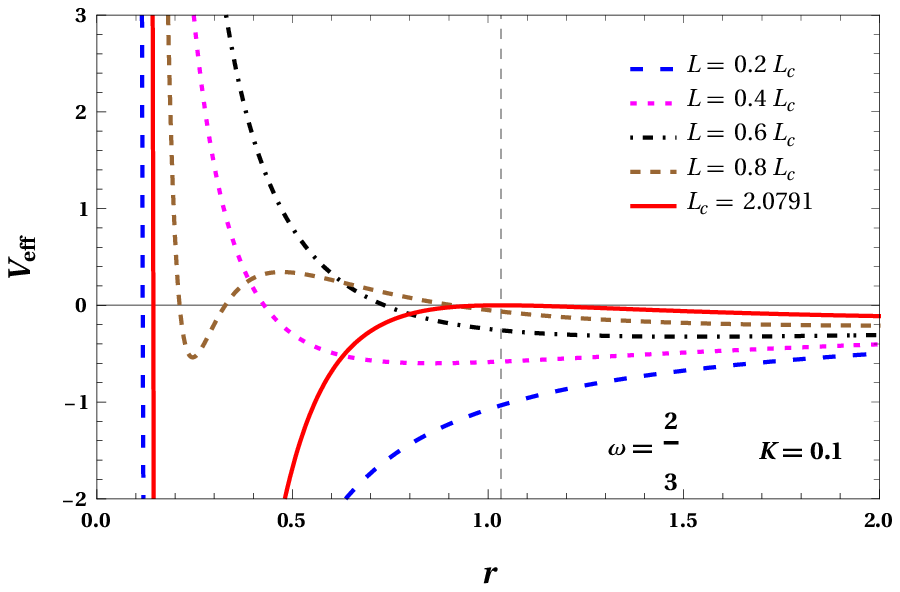}\label{kmveff3a}}
\subfigure[][]{\includegraphics[scale=0.8]{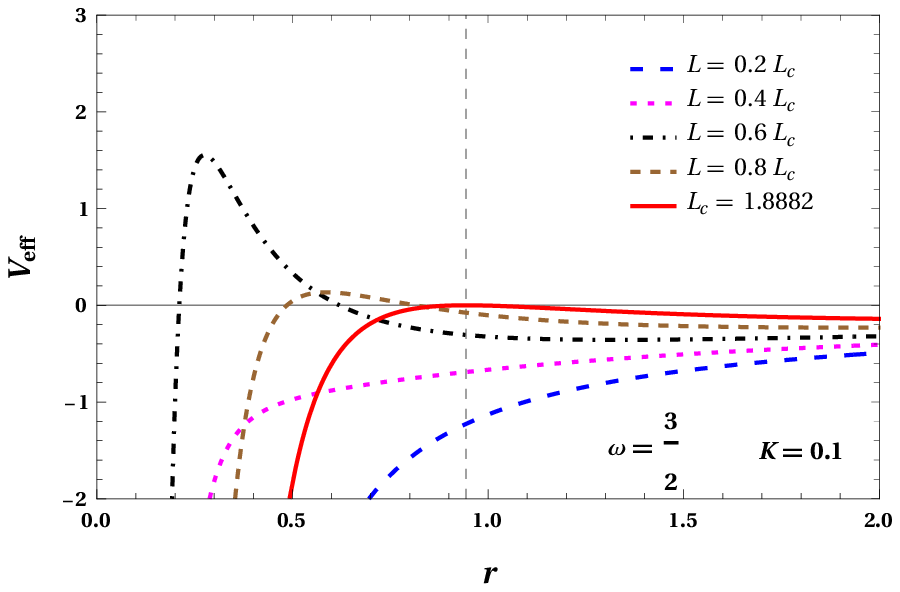}\label{kmveff4a}}

\caption{The behaviour of effective potential $V_{eff}$ below the critical angular momentum for a rotating black hole with an anisotropic matter field for different values of $(w , k)$. In all cases, for a particle approaching the black hole has no potential barrier outside the event horizon. The plots are depicted by taking the corresponding $a_e$ values.}
\label{Veffective1}
\end{figure}

\begin{figure}[tbp]
\centering
\subfigure[][]{\includegraphics[scale=0.8]{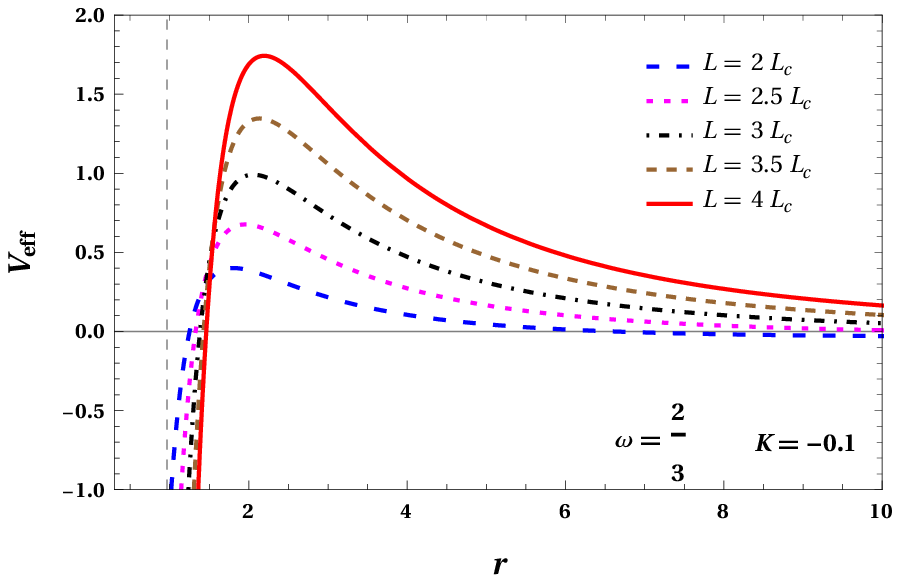}\label{kmveff1b}}
\subfigure[][]{\includegraphics[scale=0.8]{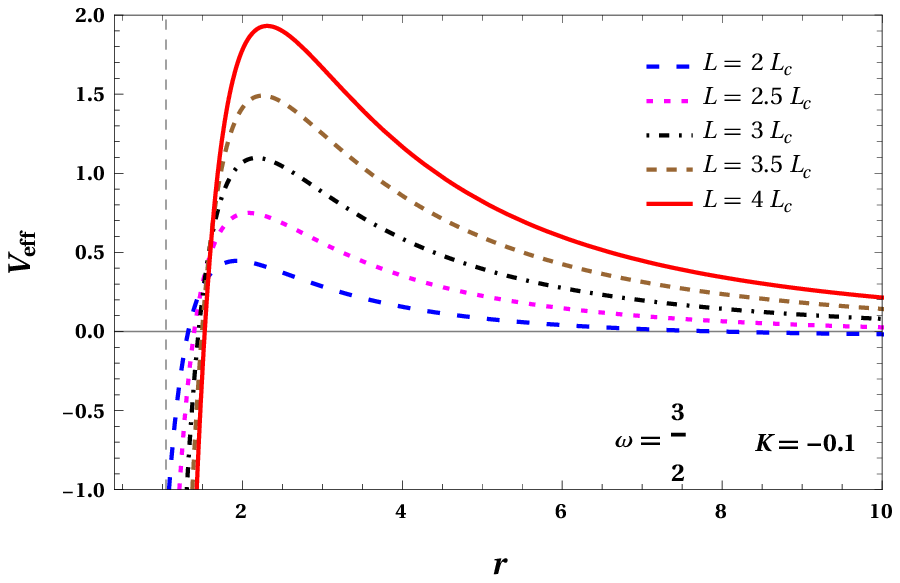}\label{kmveff2b}}

\subfigure[][]{\includegraphics[scale=0.8]{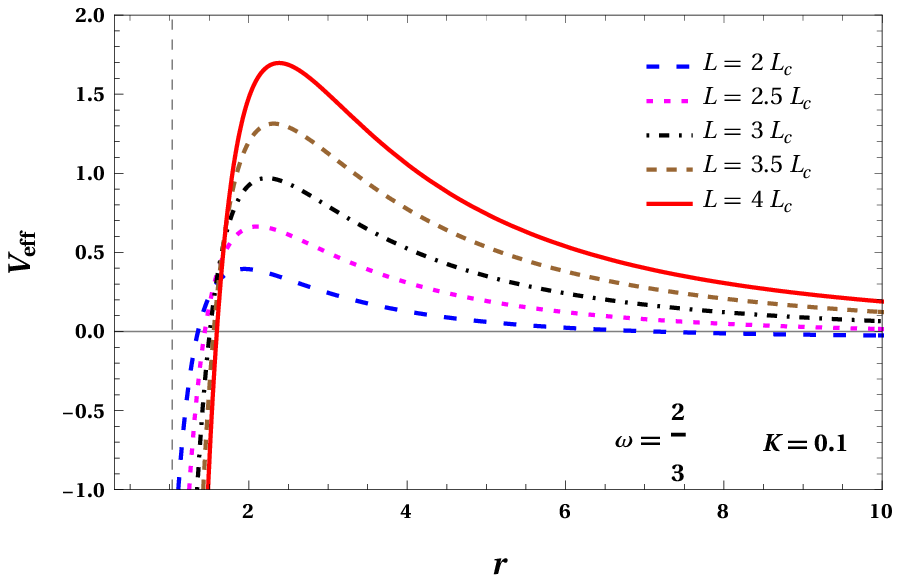}\label{kmveff3b}}
\subfigure[][]{\includegraphics[scale=0.8]{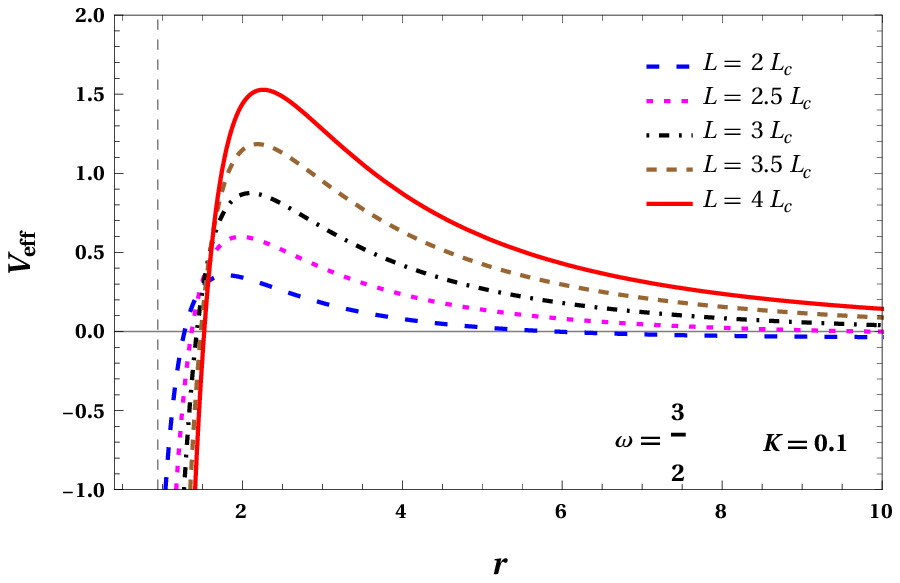}\label{kmveff4b}}

\caption{The behaviour of effective potential $V_{eff}$ above the critical angular momentum for a rotating black hole with an anisotropic matter field for different values of $(w , k)$. In all these cases, particle cannot reach the event horizon as it encounters a potential barrier. The plots are depicted by taking the corresponding $a_e$ values.}
\label{Veffective2}
\end{figure}

\begingroup
\setlength{\tabcolsep}{9pt} 
\renewcommand{\arraystretch}{1.2} 
\begin{center}
\begin{table}
\centering
\begin{tabular}{ |c|c|c|c|c|c|c|c|c| } 
\hline
&&&&\multicolumn{2}{|c|}{massive particle}&\multicolumn{2}{|c|}{photon}\\
 \cline{5-8}
$(w , k)$ & $a$ & $r_H^-$&$r_H^+$ &$L_4$(min)&$L_3$(max)&$L_4$(min)&$L_3$(max)\\
\hline
\hline
 $(2/3, -0.1)$ & 0.5 & 0.327315 & 1.60254 & -4.21892 & 3.05085 & -5.76201 & 3.54781 \\
 $(3/2, -0.1)$ & 0.5 & 0.465835 & 1.66323 & -4.33436 & 3.18442 & -5.92114 & 3.71522 \\
 $(2/3, 0.1)$ & 0.5 & 0.264345 & 1.80514 & -4.46571 & 3.37788 & -6.1239 & 4.00614 \\
 $(3/2, 0.1)$ & 0.5 & - & 1.7465 & -4.35298 & 3.25164 & -5.96877 & 3.84653 \\
\hline
\end{tabular}
\caption{\label{lrangenonext} The range of angular momentum of the infalling particle (photon or massive particle) for the non-extremal rotating black hole with an anisotropic matter field.}
\end{table}
\end{center}
\endgroup
Since the geodesics are time like, $dt/d\tau >0$. From equation \ref{teqn} we get,
\begin{equation}
\frac{1}{r^2}\left[ a(L-aE)+\frac{r^2+a^2}{\Delta}P(r) \right] \geq 0.
\end{equation}
This, under the limit $r-\rightarrow r_H^E$, reduces to
\begin{equation}
E-\Omega _H L\geq 0,
\end{equation}
where $\Omega _H$ is the angular velocity of the black hole on the horizon,
\begin{equation}
\Omega _H=\frac{a}{r_H^E+a^2}.
\label{omegaeqn}
\end{equation}
This gives the critical angular momentum of the particle to be  $L_c = E/{\Omega _H}$. For angular momentum $L$ greater than the critical values, the particle will be scattered off before reaching the horizon. Whereas, for angular momenta less than the critical values the particle is always guaranteed to cross the horizon and get absorbed by the black hole. These behaviours can easily be understood from $V_{eff}$ plots. For all $L<L_c$, effective potential is always negative which results in a bounded motion (fig. \ref{Veffective1}). Whereas, for an angular momentum less than the critical value, the particle experiences an effective potential barrier as it approaches the black hole which leads to an unbounded motion ( fig. \ref{Veffective2}). This behaviour remains same for all valid values of $(w ,K)$. In some cases there are potential bumps present for $L<L_c$ inside the horizon, which does not affect the particle dynamics outside the horizon.

\section{Centre-of-mass energy in the background of a rotating black hole with matter field}
\label{secthree}
\begin{figure}[tbp]
\centering
\subfigure[][]{\includegraphics[scale=0.8]{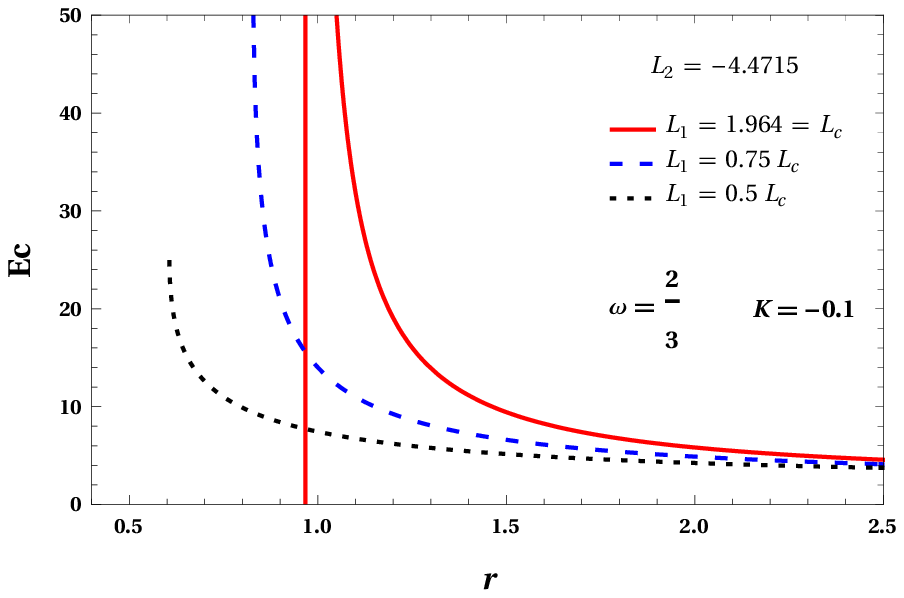}\label{kmecm1a}}
\subfigure[][]{\includegraphics[scale=0.8]{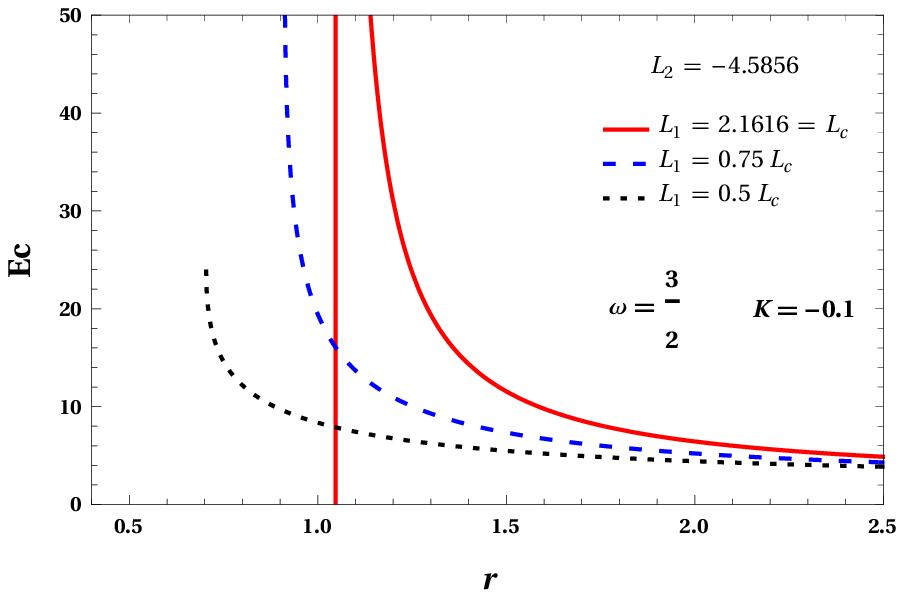}\label{kmecm2a}}

\subfigure[][]{\includegraphics[scale=0.8]{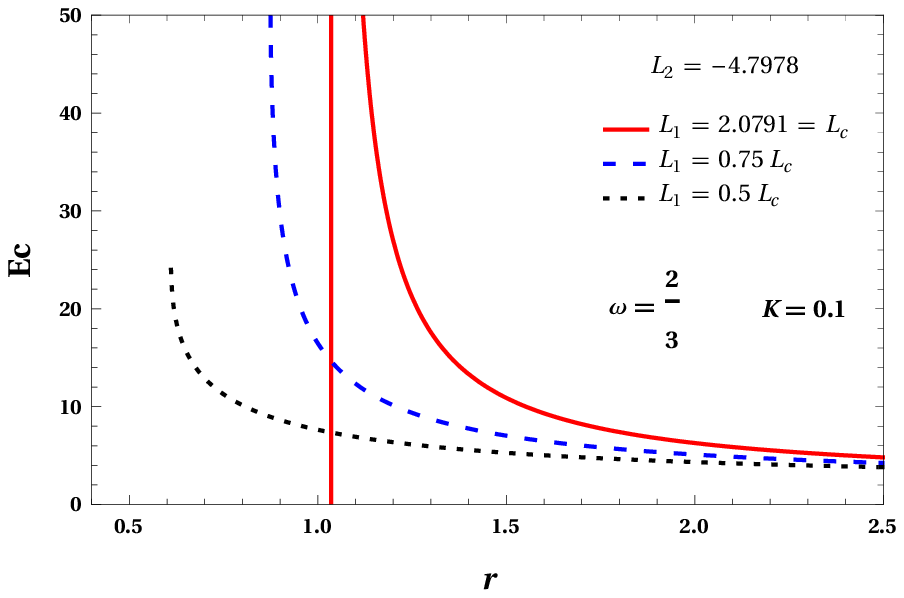}\label{kmecm3a}}
\subfigure[][]{\includegraphics[scale=0.8]{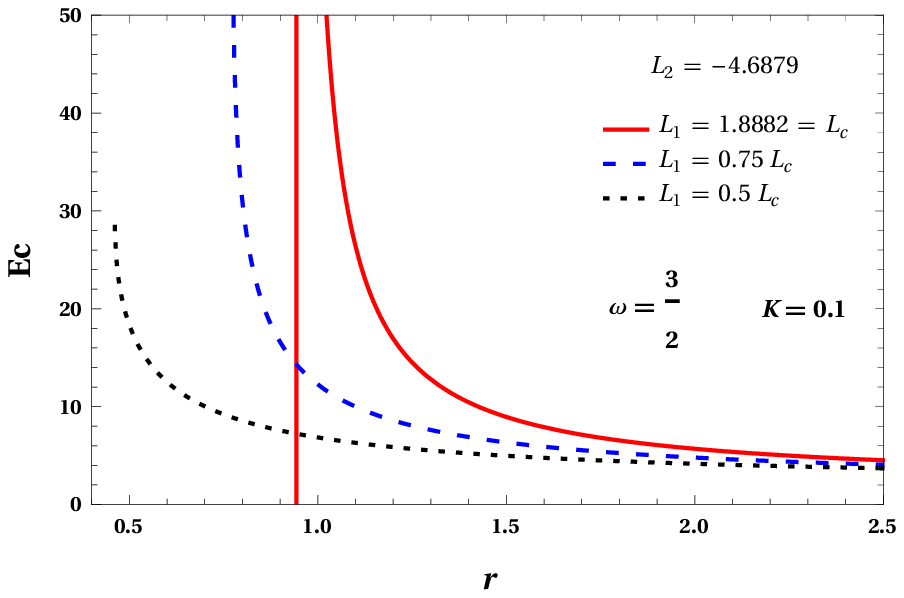}\label{kmecm4a}}

\caption{The behaviour of centre-of-mass energy for extremal black hole for different $(w , k)$ values. In all cases, the corresponding $a_e$ and $r_H^e$ values are taken. The vertical line represents the event horizon. The centre-of-mass energy diverges (red solid line) for cases where one of the particle approaches with critical angular momentum. Without lose of generality, we have taken infalling particles as general massive particles.}
\label{Ecomext}
\end{figure}

\begin{figure}[tbp]
\centering
\subfigure[][]{\includegraphics[scale=0.8]{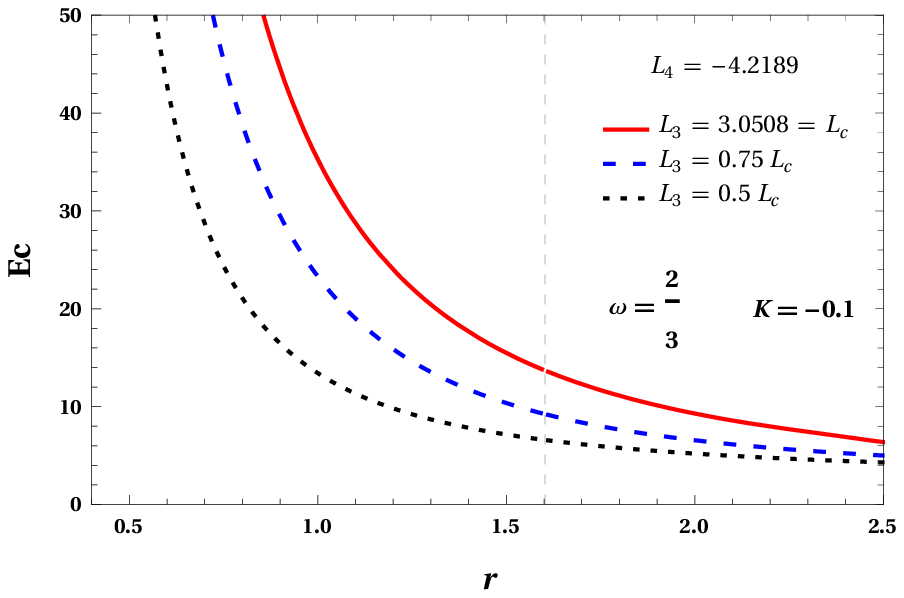}\label{kmecm1ab}}
\subfigure[][]{\includegraphics[scale=0.8]{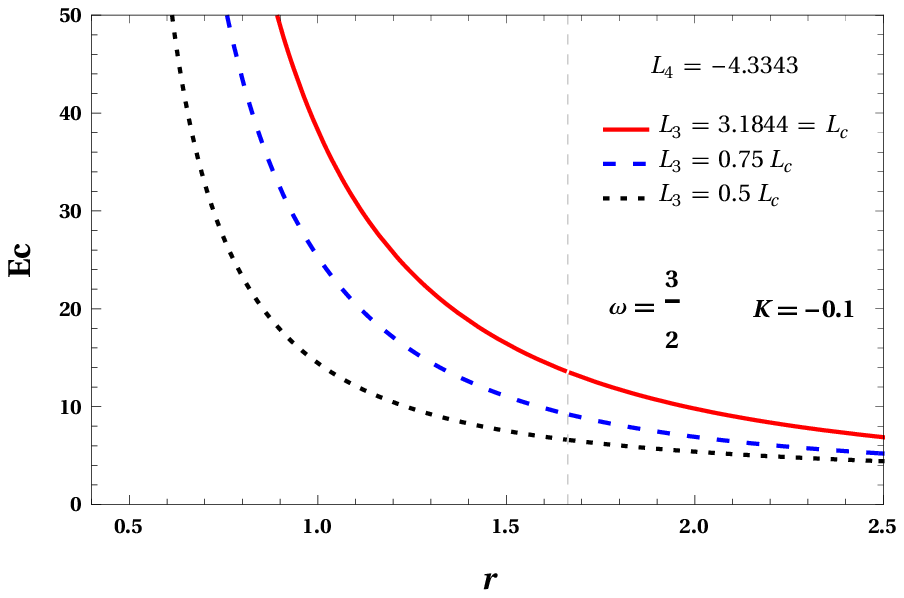}\label{kmecm2b}}

\subfigure[][]{\includegraphics[scale=0.8]{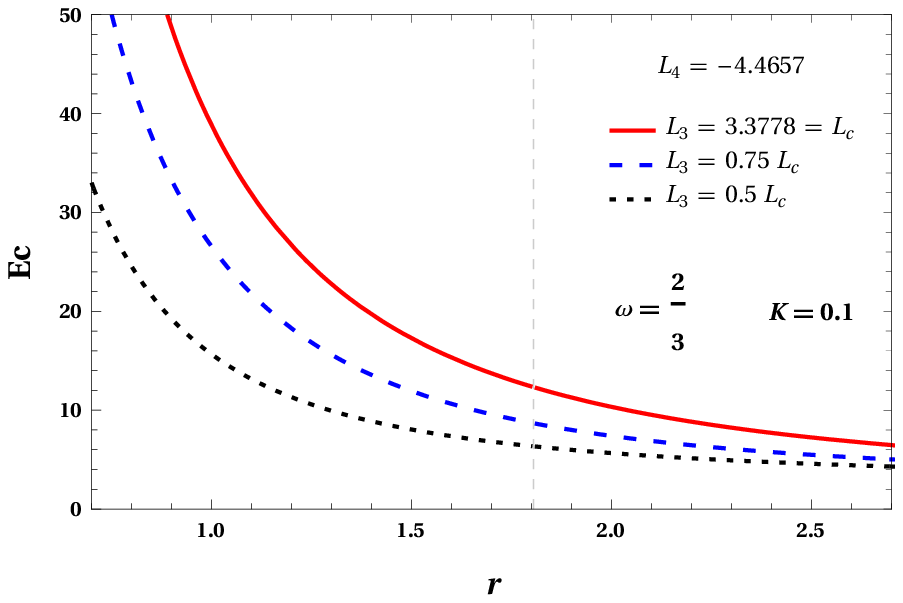}\label{kmecm3b}}
\subfigure[][]{\includegraphics[scale=0.8]{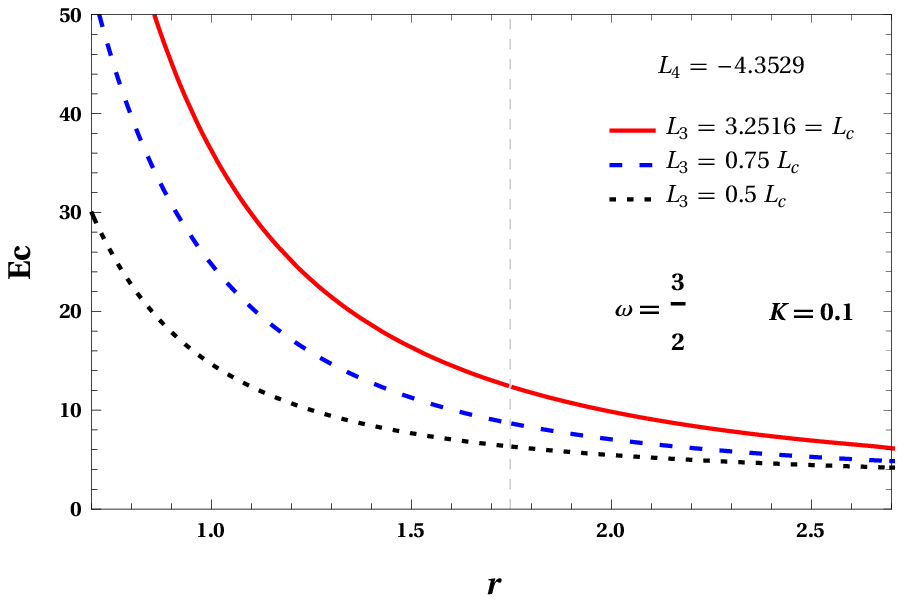}\label{kmecm4b}}

\caption{The behaviour of centre-of-mass energy for non extremal black hole for different $(w , k)$ values, $\alpha =0.005$ (left) and $\alpha =0.01$ (right). In all cases, the corresponding $a_e$ and $r_H^e$ values are taken. The vertical dotted line represents the event horizon. For all allowed values of angular momentum of the incoming particle, the centre of mass energy takes finite value outside the horizon. Without loss of generality, we have taken infalling particles as general massive particles.}

\label{Ecomnonext}
\end{figure}

Here, we study the centre of mass energy for the collision of two particles moving on the equatorial plane of a rotating black hole with matter field. We consider two uncharged particles with the same rest mass $\mu$, which are at rest at infinity initially. They approach the black hole and collide at a distance $r$ from the black hole. We assume that these particles have different angular momenta $L_1$ and $L_2$. For the collision to take place in the neighbourhood of horizon, the values of $L_1$ and $L_2$ must lie within the range of angular momentum we calculated in the previous section. As the particles are moving in a curved spacetime background, the energy in the centre of mass frame should be calculated as \cite{Banados:2009pr},
\begin{equation}
E_{CM}=\sqrt{2}\mu \sqrt{1-g_{\mu \nu} u^\mu _1 u^\nu _2}
\label{ecomeqn}
\end{equation}
where $u_i^\mu=dx_i^\mu/d\tau$ ($i=1,2$) are the four velocities of the two particles, which can be easily identified from equations of motion.  For an uncharged particle moving in the equatorial plane we have,
\begin{equation}
    u^{\mu} _i=\left( \frac{a(L_i-aE_i)}{r ^2}+\frac{r^2+a^2}{r ^2 \Delta }P(L_i) , \frac{\sqrt{\mathcal{R}}}{r^2}, 0, \frac{(L_i-aE_i) }{r ^2 }+\frac{a}{r ^2 \Delta}P(L_i)\right)
\end{equation}
Substituting this in Eq. \ref{ecomeqn}, and taking $E=1$, we obtain,

\begin{eqnarray}
E^2_{CM}=\frac{2\mu ^2}{\Delta  r^2} \left[ (a^2+r^2-\Delta)[r^2+(a-L_1)(a-L_2)-r^2L_1L_2+X_1X_2] \right],
\end{eqnarray}
where
\begin{equation}
X_i=\sqrt{\left(a L_i-r^2-a^2\right)^2-\Delta  \left((L_i-a)^2+\mu ^2 r^2\right)} \quad (i=1,2).
\end{equation}
It is clear that the centre of mass energy $E_{CM}$ is invariant under the action $L_1 \leftrightarrow L_2$, which is inevitable as we have taken identical particles with different angular momenta. As $\Delta$ depends on $(w , K)$, $E_{CM}$ is accordingly influenced by those parameters. This confirms our intuition on the deviation of the particle acceleration results for rotating black hole with matter field from that of Kerr and Kerr-Newman black holes. In addition to the influence by the surrounding matter field,  the other black hole parameters $Q$ and $a$ also affect these results, which appears through $\Delta$ only. We study the properties of centre of mass energy as the radial coordinate $r$ approaches the event horizon $r_+$ of the black hole. The results are shown in fig. \ref{Ecomext} and fig. \ref{Ecomnonext} for different combinations of $(w, K)$. In each case we have taken the angular momentum of the in-going particles within the allowed range to ensure the collision to be in the vicinity of the horizon. The critical angular momentum, calculated using Eq. \ref{omegaeqn}, lies outside the range $(L_{min}, L_{max})$ for non extremal black holes, whereas, it is within the range for extremal black holes. Only a particle with critical angular momentum will reach the horizon.

From the plots it is clear that the value of $E_{cm}$ is indeed finite for generic values of $L_1$ and $L_2$. However, for extremal black hole, the centre-of-mass energy $E_{CM}$ is unlimited when one of the particle has a critical angular momentum. Arbitrarily high CM energy near the horizon is interesting, as  the result may provide a possible way to probe the Planck-scale physics in the background of a rotating black hole with matter field. Compared to the cases of Kerr and Kerr-Newman black holes, the spin of the black hole with matter field can be greater than unity, implying a enhanced centre of mass energy. The scenario with the extremal black hole is an idealized one, as it takes infinite timefor a particle to reach the horizon from infinity. However the flying time is finite for non extremal black holes. Therefore we expect the near-extremal cases can act as real particle accelerators, even though the time taken is very large. We investigate behaviour of $E_{cm}$ for non extremal black holes in fig. \ref{Ecomnonext}, where it has finite and sufficiently high values at the horizon for all cases.



\section{Discussions}
\label{secfour}
In this article we have demonstrated that a rotating black hole with an anisotropic matter field can act as a particle accelerator. Such a black hole solution is interesting from astrophysical point of view, as stellar objects are rotating and surrounded by fluids or fields. The properties deviate from that of Kerr and Kerr-Newman solutions due to the density and anisotropy of the surrounding matter field, which described by the parameters $K$ and $w$, respectively. However the choice of $(w, K)$ must satisfy appropriate physical conditions like energy conditions. For the allowed values of $(w, K)$ we have analysed the horizon structure and ergo sphere, which are significantly modified in comparison to Kerr and Kerr Newman solutions. An important point in this regard is that the black hole can have spin greater than Kerr-Newman case, which is facilitated by the anisotropic fluid. For a given value of $(w, K)$ we have extremal and non extremal black holes, depending on the values of $Q$ and $a$. For a fixed charge $Q$, we can have the extremal case with spin $a_e$.

By using the BSW mechanism, we analysed the properties of the centre-of-mass energy for two particles colliding in the equatorial plane of the black hole. The particle motion is analysed by solving the geodesic equations, from which we obtain the range of angular momentum for which the particle reaches the horizon of the black hole. This in fact is a manifestation of the effective potential to offer a window or barrier for the incoming particle. For the particles within the allowed window, we have studied the particle collision near the black hole. It is observed that, when one of the particles is arriving with critical angular momentum, the centre of mass energy is arbitrarily high in the space time of the extremal black hole. However, the energy is finite for the case of non-extremal black holes, even if the ingoing particle has critical angular momentum. As the horizon structure is determined by the anistropic matter field parameters $(w,K)$, the BSW mechanism depends on these values.

The demonstration that centre-of-mass energy diverges in particle collision, is interesting in the context of astrophysics. Super massive black holes can accelerate particles in this mechanism, which can be related to the observed ultra high energy cosmic rays up to $10^{20} eV$. A possible by product of these particle collisions is, exotic massive particles, which can also be analysed in a realistic scenario with our results. Therefore,  we hope that our study can be related to the observational aspects to probe Planck scale physics.


\acknowledgments
Author A.R.C.L.,  N.K.A. and K.H. would like to thank U.G.C. Govt. of India for financial assistance under UGC-NET-SRF scheme.


  \bibliography{BibTex}

\end{document}